\documentclass[a4paper,11pt]{amsart}
\begin{document}

\title{{\bf  THE SUPERMASSIVE CENTRE OF OUR GALAXY}
\\*\emph{ET CETERA}}
\author{Angelo Loinger}
\date{}
\address{Dipartimento di Fisica, Universit\`a di Milano, Via
Celoria, 16 - 20133 Milano (Italy)}
\email{angelo.loinger@mi.infn.it}
\author{Tiziana Marsico}
\date{}
\address{Liceo Classico ``G. Berchet'', Via della Commenda, 26 - 20122
                                           Milano (Italy)}
\email{martiz64@libero.it}
\thanks{to be published on \emph{Spacetime \& Substance}}

\maketitle

\begin{abstract}
We show that the supermassive celestial body at the centre of the Milky Way
and the two supermassive celestial bodies at the centre of the distant galaxy NGC 6240
cannot be black holes.
\end{abstract}

\vskip1.20cm
%\section{Introduction}
\noindent{\bf  Introduction}\par \vskip0.10cm
Sophisticated observations made by a team of astrophysicists of the M.P.I.
for Extraterrestrial Physics have allowed to determine the positions of the
star denoted with the symbol S2 in its motion around the Milky-Way's centre \cite{1}.
It has come out that the S2 orbit is a Keplerian ellipse with a period of  15.2  years.
\par
We acknowledge the accuracy of the above research, but we are
rather sceptical about the conclusion of the authors according to
which the centre around which S2 revolves is a black hole. As far
as an ``explanation'' is an explanation, it is necessary that what
is explained is a \emph{logical} consequence of the premiss and of
what is used for explaining. As a matter of fact, the existence of
the observed Keplerian orbit can only explain the presence of a
punctual supermassive body at the centre of the Milky Way -- and
not of a supermassive black hole. This conclusion will be
corroborated in a detailed way in the following sections.

\vskip0.80cm
%7section{}
\noindent {\bf 1.}-- {\bf  Theoretical considerations} \par \vskip0.10cm
If $r, \theta, \phi$ are spherical polar coordinates, the solution
of the problem of the Einsteinian gravitational field generated by
a point mass $M$ at rest is given by the following spacetime interval \cite{2}:
$$
{\rm d}s^2=[1-\frac{2m}{f(r)}]c^2{\rm d}t^2-[1-
\frac{2m}{f(r)}]^{-1}[{\rm d}f(r)]^2-f^2(r)[{\rm d}\theta^2+\sin^2\theta {\rm d}\phi^2], \quad  (1)
$$
where: $m\equiv MG/c^2$; $G$ is the gravitational constant and
$c$ the speed of light  \emph{in vacuo}; $f(r)$ is \emph{any} regular
function of $r$.
\par
If one chooses $f(r)\equiv r$, one obtains the so-called standard
form of solution, \emph{erroneously} named ``by Schwarzschild'',
but in reality due to Hilbert, Droste and Weyl \cite{2}.
\par
One has the \emph{original} form of solution given by
Schwarzschild in 1916 if one chooses $f(r)\equiv
[r^3+(2m)^3]^{1/3}$; Schwarzschild's ${\rm d}s^2$ holds in the
\emph{entire} spacetime, with the only exception of the origin
$r=0$: it is ``maximally extended''. Remark that Schwarzschild's
form of solution is diffeomorphic to the part $r>2m$ of the
standard form. For $r<2m$ this part loses any mathematical and
physical meaning - as it was repeateadly emphasized by Einstein
and by all the Great Men who developed the general relativity --
because the solution becomes non-static, the radial coordinate $r$
becomes a time coordinates, and the ${\rm d}s^2$ loses its
physical ``appropriateness''. Now, the invention of the senseless
notion of  black hole was originated by an odd reflection on the
region $r<2m$. If the treatises had expounded the original form of
Schwarzschild in lieu of the standard form, the notion of black
hole would not have come forth.
\par
In conclusion, the physical results are those and \emph{only} those that
are \emph{independent of the particular choice} of the function $f(r)$.
But the fictive notion of black hole owes its origin to a misinterpretation
of the part $r<2m$ of a \emph{particular} form: the \emph{standard} form.

\vskip0.80cm
%\section{}
\noindent {\bf 2.}-- {\bf Experimental results and conclusions}
\par \vskip0.10cm Some months ago Sch\"odel and other 22 authors
published a paper \cite{1} in which they report ``ten years of
high-resolution astrometric imaging'' that have allowed ``to trace
two-thirds of the orbit of the star, [denoted with S2], currently
closest to the compact radio-source Sagittarius A$^\ast$
[SgrA$^\ast$].'' They write: ``The observations, which include
both pericentre and apocentre passages, show that the star is on a
bound, highly elliptical keplerian orbit around SgrA$^\ast$, with
an orbital period of 15.2 years and a pericentre distance of only
17  light hours. The orbit with the best fit to the observations
requires a central point mass of  $(3.7 \pm 1.5)\times 10^6$ solar
masses ($M_\odot$). The data no longer allow for a central mass
composed of a dense cluster of dark stellar objects or a ball of
massive, degenerate fermions.''
\par
Sch\"odel \emph{et alii} \cite{1} interpret their results as an experimental proof of
the existence of a supermassive black hole at the centre of our galaxy.
Evidently, they are victims of the very diffuse, but erroneous opinion
(see sect.{\bf 1}) that the gravitational collapse of a massive celestial body
must generate a black hole. On the other hand, the experimentally
observed orbit is \emph{Keplerian},  that is described by the (nonrelativistic)
Newtonian theory. From the standpoint of \emph{logic and experience},
one can \emph{only} affirm that at the centre of Milky Way there is a
punctual object with a huge mass. (It is interesting to remark that
the fictitious event horizon for a point mass $\approx 4\times 10^6 M_\odot$
would be situated at a distance of $\approx 26$ light seconds!).

\vskip0.80cm
%\section{}
\noindent {\bf APPENDIX} \par \vskip0.10cm In a \emph{NASA Press
Release}, dated November 20th, 2002, entitled ``A Super Galactic
Discovery'', we read: ``For the first time, scientists have found
proof of two supermassive black holes together in the same galaxy.
These black holes are orbiting each other and will merge several
hundred million years from now. The event will unleash intense
radiation and gravitational waves [....] and leave behind an even
larger black hole than before.
\par
NASA's Chandra X-ray Observatory spotted the two black holes in
the galaxy NGC 6240. The observatory was able to ``see'' them
because the black holes are surrounded by hot swirling vortices af
matter called accretion disks. Such disks are strong sources of
X-rays.''
\par
This is pure science fiction! Demostration. \emph{In primis}, as
we have seen in sect.{\bf 1}, the very notion of black hole is a
nonsense. \emph{In secundis}, even if we believed in the existence
of the black holes, there would be ``no way of asserting through
some analogy with Newtonian gravitational theory that a black hole
could be a component of a close binary system or that two black
holes could collide.  An existence theorem could first be needed
to show that Einstein's field equations contained solutions which
described such configurations.'' \cite{3}. \emph{In tertiis}, the
eventual formation of an accretion disk, strong source of X-rays,
is \emph{not} linked to a particular choice of the arbitrary
function $f(r)$ (see sect.{\bf 1}) -- in particular to the
standard choice $f(r)\equiv r$.
\par
In conclusion, no black hole has been detected by Chandra X-ray
observatory. And never gravitational waves will travel over the
world, because they are pure mathematical undulations, completely
devoid of any \emph{physical} reality \cite{2},\cite{4},\cite{5}.

 \small \vskip0.5cm\par\hfill{``\emph{La v\'erit\'e,
l'\^apre v\'erit\'e.''} \vskip0.10cm\par\hfill\emph{Danton}}

\end{document}